\documentclass[letterpaper,twocolumn,10pt,keeplastbox]{article}
\usepackage{usenix2019_v3}

\usepackage{url}
\usepackage{xspace}
\usepackage{graphicx}

\newcommand{\eg}{\textit{e.g.}\xspace,\xspace}
\newcommand{\etal}{\textit{et~al.}\@\xspace}
\newcommand{\etc}{\textit{etc.}\@\xspace}
\newcommand{\ie}{\textit{i.e.}\xspace,\xspace}

\begin{document}

\title{Automated Discovery for Emulytics}

\author{
{\rm Jonathan Crussell, David Fritz, Vince Urias}\\
Sandia National Laboratories \\
\{jcrusse, djfritz, veuria\}@sandia.gov
} 

\maketitle

\begin{abstract}
Sandia has an extensive background in cybersecurity research and is currently
extending its state-of-the-art modeling via emulation capability. However, a
key part of Sandia's modeling methodology is the discovery and specification of
the information-system under study, and the ability to recreate that
specification with the highest fidelity possible in order to extrapolate
meaningful results.

This work details a method to conduct information system discovery and develop
tools to enable the creation of high-fidelity emulation models that can be used
to enable assessment of our infrastructure information system security posture
and potential system impacts that could result from cyber threats. The outcome
are a set of tools and techniques to go from network discovery of operational
systems to emulating complex systems.

As a concrete usecase, we have applied these tools and techniques at
Supercomputing 2016 to model SCinet, the world's largest research network. This
model includes five routers and nearly 10,000 endpoints which we have launched
in our emulation platform.

\end{abstract}

\section{Introduction}

Our Nation's critical infrastructure depends on distributed information
systems, which must be resilient against malicious attempts to disrupt their
operation. Security mechanisms must be deployed and continuously assessed in
order to prevent adversaries from disrupting critical infrastructure. These
information systems employ a broad range of technologies including fixed
Ethernet, point-to-point wireless, and cellular communications. Additionally,
these systems employ enterprise architectures, cloud based data centers, and
emergency response wireless architectures. With the extensive reliance of
critical infrastructure on secure information systems, techniques and
mechanisms to assess security must be created.

Assessing the security posture of information systems used in critical
infrastructures necessitates, in cases, the capability to assess security on a
representation of the operational system. Performing security analysis on real
operational systems, in most cases, is prohibitive and assessing security on
simulated systems lacks fidelity. A cost-effective solution is to create an
emulation of the system and its components to enable system-level analysis. An
emulated system model can be used to conduct red teams and assess risks and
vulnerabilities of a system. A major challenge in creating an emulated system
model is obtaining an accurate specification of the system of interest.

Initially security practitioners may examine original system design and
specification documents, if available. This typically results in a poor
description and specification of the system since the system has been modified
for several possible reasons:

\begin{itemize}
    \item Original specification was modified during original deployment
    \item Device configurations not completely specified and/or modified over time
    \item Original device firmware and/or software upgraded
    \item Original system topology modified for system growth
    \item Device selection changes resulting from vendor performance
        improvements
\end{itemize}

In order to obtain an up-to-date and accurate view of the information system a
dynamic system discovery and mapping capability must be developed and employed.
The capability developed here enables system analysts to diagram, inventory,
audit, and analyze the system of interest. Our solution encompasses two areas:
System Discovery and System Emulation.

We have developed an information system discovery platform that can be used as
a centralized discovery station. The platform accounts both for application
(and service representation) as well as network configuration. System discovery
in critical infrastructure systems can pose a significant challenge because of
the diversity of devices not normally used in traditional corporate IT systems.
System devices may include embedded devices that do not respond to traditional
discovery techniques and discovery might be limited to network protocol
scanning mechanisms.

Our developed capability fuses data from both commercial and open-source
solutions. We can employ both active device discovery and mapping techniques
along with passive discovery techniques to discover and create maps of
information systems of interest.

\begin{figure*}[ht]
    \centering
    \includegraphics[width=0.6\textwidth]{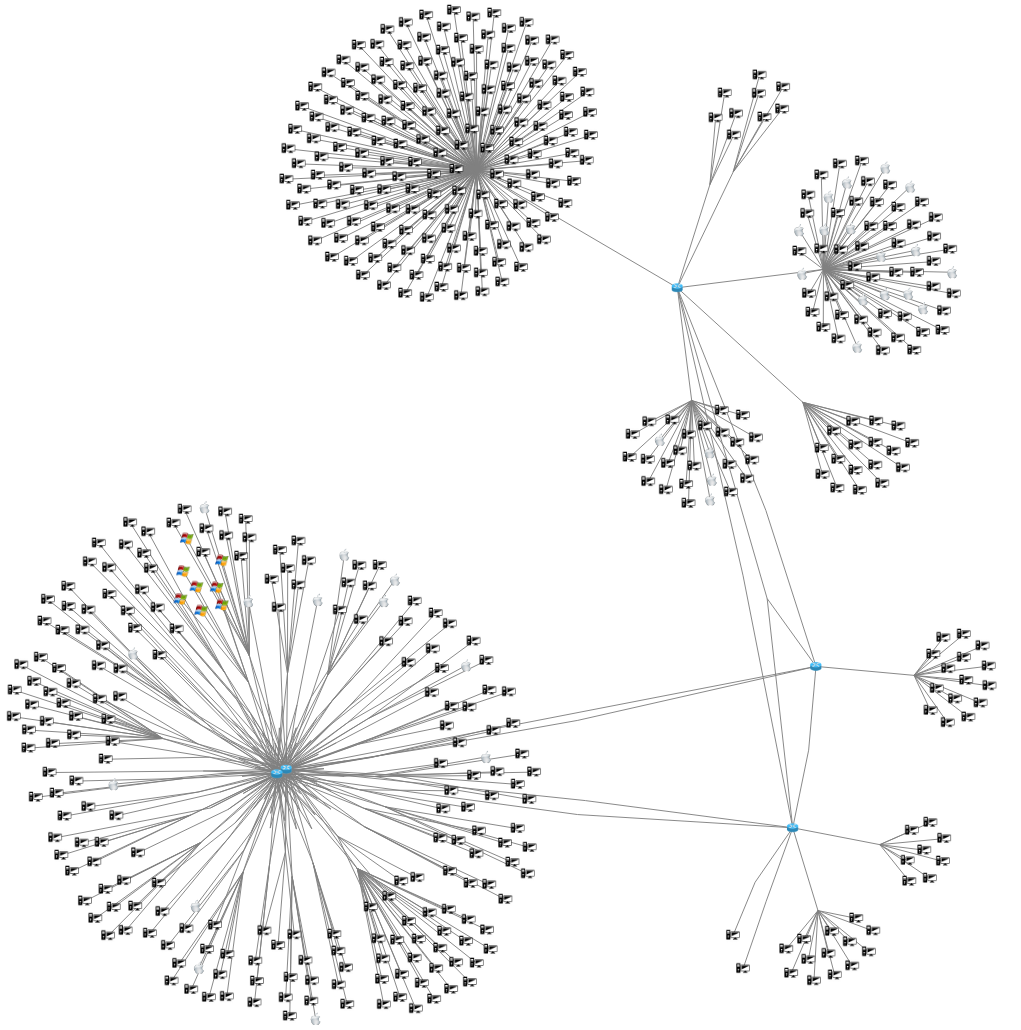}
    \caption{Model of SCinet 2016, omitting the two largest subnets: the
    wireless network and the classrooms. The aggregation routers in the bottom
    left connect all the vendor machines to the conference network.}
    \label{fig:sc16-partial}
\end{figure*}

The contributions of this work are as follows: we present the \emph{discovery}
toolset\footnote{Available at
\url{https://github.com/sandia-minimega/discovery}} and associated intermediate
representation (IR) as a means to automate the map-to-model process and apply
the toolset to the Supercomputing 2016 network (SCinet) to demonstrate its
utility (shown in Figure~\ref{fig:sc16-partial}).

\section{Background}

Sandia has a long history in large-scale network
emulation~\cite{urias2015emulytics} using virtual machines (VMs). Specifically,
Sandia has developed \emph{minimega}\footnote{\url{http://minimega.org}}, an
open-source VM orchestration tool. minimega supports Sandia's
Emulytics\footnote{\url{https://www.sandia.gov/emulytics/}} program that
combines large-scale network emulation with data analytics.

minimega is fast, easy to deploy, and can scale to run on massive clusters with
virtually no setup. minimega uses
QEMU/KVM~\cite{bellard2005qemu,habib2008virtualization} to launch VMs (which
can run unmodified Windows and Linux operating system). These VMs can be
configured to have different resource allocations (\eg VCPUs and memory) and
properties (\eg disk images, number of serial ports). For higher density,
minimega also includes support for lightweight virtualization using Linux
containers. Importantly, minimega uses Open vSwitch~\cite{pfaff2015design} to
connect these VMs and containers using virtual switches to create arbitrary
network topologies according to the user's specifications.

Once configured, minimega includes many capabilities to launch and manage the
VMs and containers (simply referred to as VMs here). minimega has a scheduler
that launches the VMs across a cluster of machines to evenly distribute
resource allocations. Once launched, minimega has APIs to start, pause, and
kill VMs as well as inspect their current state. minimega also has a
command-and-control system that talks to agents running within the VMs to send
and receive files and execute commands. For example, this can be used to start
traffic generation and then pull the log files for analysis. Additionally,
minimega has APIs to configure and deploy routers based on
BIRD~\cite{filip2010bird}. These routers provide routing, DNS, and DHCP for the
VMs. Finally, minimega has APIs to record traffic from the VM networks for
further analysis.

These capabilities allow experimenters to build complex networks of VMs and
conduct tests with minimega. However, the output quality of these emulations is
bounded by the quality of the input system specification.

\section{Related Work}

Many tools have been built to date for network discovery. Tools can generally
be divided into two categories: topology and endpoint discovery. Topology
discovery tools try to determine the structure of the network while endpoint
discovery tools provide detailed information about individual endpoints. These
tools can be further categorized into active tools, when the tool is allowed to
probe the network, and passive tools, when they are not.

Traceroute~\cite{jacobson1988traceroute} is an example of an active tool that
has been used to infer network topologies. Traceroute enumerates the routing
path between a source and destination. Many techniques have been developed on
top of traceroute to construct a topology from the
paths~\cite{govindan2000heuristics,kim2007efficient,spring2002measuring}.
Nmap~\cite{lyon2009nmap} is an active scanning tool to enumerate properties of
endpoints. Nmap can discover operating systems, open ports, and sometimes
versions for applications listening on those open ports. In order to do so,
Nmap generates many probes that may be blocked by firewalls. There are other,
passive, tools that provide situational awareness built on top of
NetFlow~\cite{berthier2010nfsight,yin2004visflowconnect}. These tools can be
used for both topology and endpoint discovery.

These network discovery tools are all complementary to our system -- we
envision our system ingesting output from any and all these tools to build a
detailed and rich model of the network. In fact, we prefer to use these other
tools and write parsers for their outputs rather than implement the full tool
ourselves. This simplifies our development and allows us to leverage the
decades of development that have gone into these other tools.

Mirkovic \etal\cite{mirkovic2018dew} present Distributed Experiment Workflows
(DEW) which describes experiment topology as well as experiment behavior and
topological constraints. Our approach could be combined with DEW to allow
researchers to define experiments on automatically discovered topologies.

In addition to Sandia's minimega, there are other emulation platforms such as
DETER~\cite{benzel2006experience}, Emulab~\cite{hibler2008large}, and
CloudLab~\cite{ricci2014introducing}. Each of the emulation platforms have
their own set of design principles and goals and could have been used instead
of minimega. Fortunately, the majority of the discovery toolset is platform
agnostic. The final step which emits minimega commands could be replaced with a
separate tool that automates the APIs of these other platform.

Depending the application of the network model, full-system emulation may not
be necessary. In which case, tools like mininet~\cite{de2014using}, which is a
tool for software-defined network developers to prototype large networks on a
single computer using lightweight virtualization, may be appropriate.

\section{Methodology}

There are two key aspects involved in this work. First, to investigate methods
for rapid, automated, and flexible network structure and behavior specification
(raw data to map). This effort required studying issues both in data transforms
appropriate for an Emulytics environment, but also issues in composing data in
constructive ways. Second, given a suitable representation of a network, create
a mechanism to emit Emulytics models with variable levels of abstraction
suitable for operators to create, adjust, and analyze.

\subsection{Rapid, automated, flexible network specification (raw data to map)}

The overall goal of this work is to investigate mechanisms to create Emulytics
models from source data (\eg network packet capture, active network scans,
router configurations). We decouple this operation into transforming sensor
data into an intermediate representation and then transliterating the
intermediate representation into an Emulytics model. This is not unlike the
process most code compilers use -- take human readable source, transform into
an intermediate representation (parse tree), make optimizations, transform into
machine code or other representations. By creating an intermediate
representation, we create opportunities to perform analyses on the
specification such as identifying partitions in the network, as well as create
a portable and reusable specification for the model generator to create
multiple models from.

The intermediate representation (IR) is a simple graph, with unstructured
key/value pairs on each node (\ie an endpoint or switch in a network), as well
as unstructured key/value pairs on each edge (\ie a network connection). Nodes
and edges are populated with descriptive data about their role and operation on
a network (\eg IP and MAC address, running services, hostname).

We have created a number of parsers for a variety of source data types that can
interact with the data already in the graph. For example, when parsing network
packet capture data, the parser may find that a node with IP address 10.0.0.1
is communicating to a specific web server. To capture this behavioral artifact,
the parser searches the graph for a node with IP 10.0.0.1. If it finds one, it
adds this metadata to the node specification (perhaps incrementing a total byte
count of traffic as well). If the node does not exist, it simply creates one,
attaching it to the correct subnet (assuming that it can be inferred), and
moves to the next packet. Using this model, we support both creating any number
of discrete parsers and provide a path to transliterate data from existing
commercial or open source tools (by modifying those tools or creating ``shim''
parsers).

The host IR also enables us to automatically perform ``gap'' analysis before
emitting an Emulytics model. Continuing with the above example, a compiler is
able to perform certain checks on the IR of software such as ensuring that
certain semantics are met such as that variables are defined before they are
used. We envision performing similar semantic checks on the host IR. For
example, it would be syntactically valid to define a machine with thousands of
interfaces in our host IR but such a machine would be physically unrealizable
and, thus, a semantic checker of the host IR would reject it as invalid and
suggest potential alternatives.

Likewise, we could encode the best practices of networking engineers as
semantic checkers. These semantic checkers could also identify blind spots in
the network where it can identify the host IR is incomplete. Without an IR,
such checks would not be feasible.

\subsection{Tunable model specification (map to model)}

The second stage of creating an Emulytics model from source data is to generate
a model from a suitable IR that has been created, optimized, and validated. For
the purposes of this project, we assume that we have sufficient data in the
intermediate representation to create a bootable Emulytics experiment (\ie all
endpoints have IP addresses or are configured for DHCP). The primary goal of
the map-to-model stage is to support translating the IR to an Emulytics model.
This stage also allows for tuning the translation to support emitting
variations of the model. For example, the IR may be comprised of 250,000
Windows-based endpoints. This is too many Windows virtual machines to boot on
most clusters. Instead, the operator should be able to identify that only some
subsection of endpoints are important to a given experiment and need to
actually run Windows virtual machines. The rest of the model may safely be made
up of generic Linux containers, which allow for greater density, without
impacting the validity of the experiment. This specification abstraction does
not modify the source IR, simply the generated model, meaning that the operator
can revisit the intermediate representation with new model criteria for other
experiments.

\section{Implementation}

\begin{figure*}
    \centering
    \includegraphics[width=0.65\textwidth]{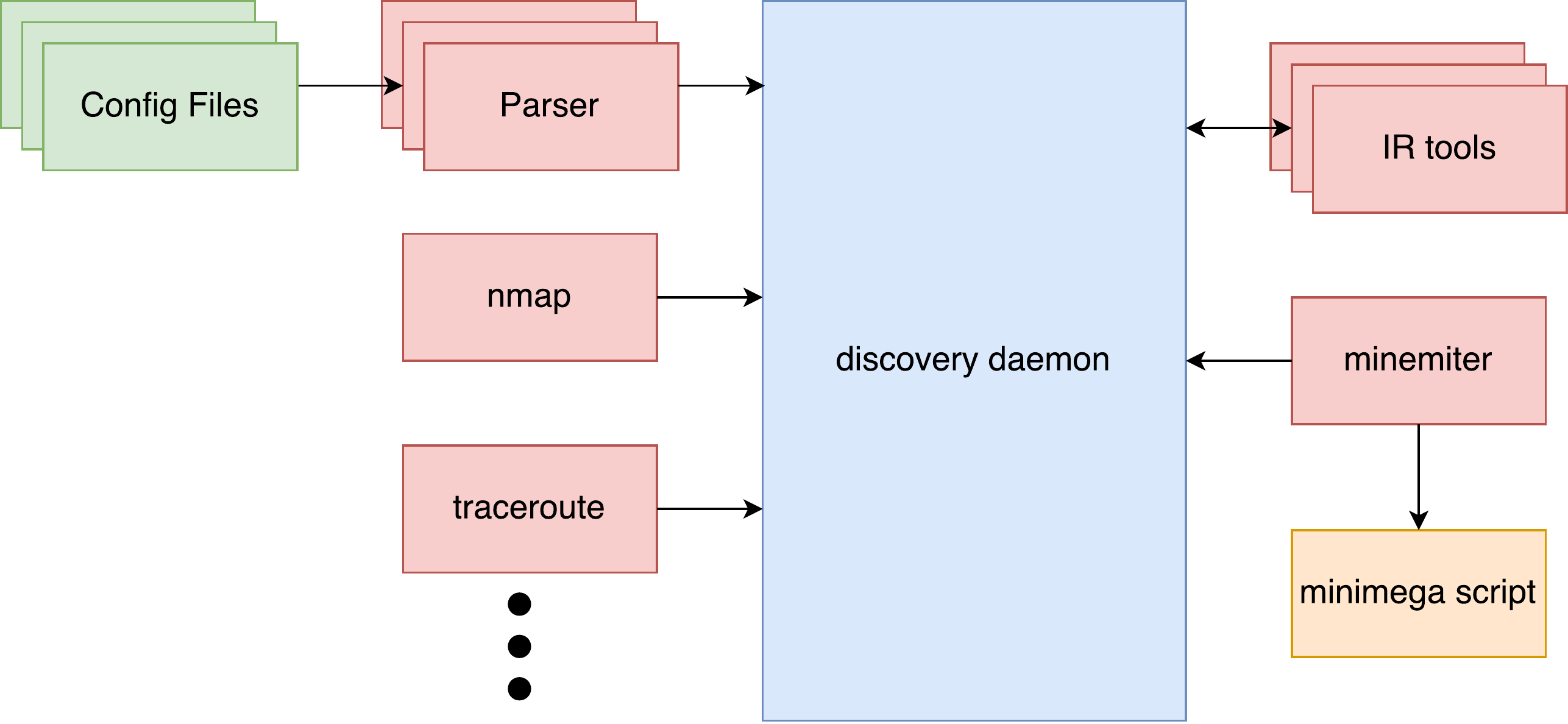}
    \caption{Overview of the discovery toolset. The discovery daemon accepts input from raw data parsers to build the IR. IR tools may then check and refine the IR before minemiter creates a minimega script.}
    \label{fig:overview}
\end{figure*}

We built the discovery toolset to integrate into the existing minimega
ecosystem. As such, all of the tools are implemented in Go. \emph{discovery} is
the main tool in the toolset. It acts as the central tool which all other
tools, parsers, model generators, \etc communicate with, and stores the IR of
experiments, as well as other configuration data. In addition to storing the
IR, the discovery tool also hosts a simple web server that provides a real-time
visualization of the experiment (as seen in Figure~\ref{fig:sc16-partial} and
Figure~\ref{fig:sc16-routers}). Figure~\ref{fig:overview} shows an overview of
the toolset's interactions.

\begin{figure}
    \begin{verbatim}
{
    "NID": 1,
    "Edges": [
        {
            "N": 63,
            "D": {
                "ip": "10.0.0.1/24",
                "mac": "de:ad:be:ef:ca:fe"
            }
        }
    ],
    "D": {
        "hostname": "irc.example.com",
        "os": "linux",
        "ports": "22,6667"
    }
}
    \end{verbatim}
    \caption{Synthetic IR for a generic Linux server that runs SSH and IRC
    servers and is connected to a single network with a static IP.}
    \label{fig:host_ir}
\end{figure}

The discovery tool maintains a single IR at runtime, and supports a simple
JSON-encoded RESTful interface for queries and updates. The tool can read and
write experiments to disk as well. Figure~\ref{fig:host_ir} shows an example of
JSON encoded data for a single node in the IR. Node and edge data is stored in
unstructured maps that are later parsed by the model generator. Each node and
network can by referenced by its ID.

The tool also maintains a simple map of configuration data, intended to be used
to store information about model specific details, such as the location of disk
images that will be used by an experiment.

\subsection{Parsers}

We developed several parsers for the output from existing tools including
\emph{nmap} and \emph{traceroute}. We also develop a new PCAP parser based on
GoPacket\footnote{\url{https://github.com/google/gopacket}}.

For \emph{nmap}, we walk the XML output from the tool and create an endpoint
for each host that nmap identifies, storing the identified operating system and
open ports in the metadata. Our implementation of this parser assumes that the
structure of the network is already known -- all routers and their connected
subnets are already represented in the IR.

For \emph{traceroute}, we parse the hop-by-hop output and create an endpoint
for each hop. This demonstrates the basic concept but has a number of
shortcomings including alias resolution and determining the IPs for
``backward'' edges. These shortcomings could be addressed by incorporating
prior work~\cite{govindan2000heuristics,kim2007efficient,spring2002measuring}.

We implement our own PCAP parser to perform passive endpoint discovery and
record the results in the IR. Specifically, we parse packets using GoPacket,
and extract information from ARP, Multicast DNS, DHCP, ICMPv4, and TCP SYN
packets.

\textbf{ARP}~\cite{rfc826} is used by endpoints to identify the hardware
address associated with a network address. The ARP Replies are recorded in the
IR, creating new endpoints if the IR does not already contain a given hardware
or network address.

\textbf{Multicast DNS}~\cite{rfc6762} is a service discovery protocol used in
local networks. By snooping on multicast DNS packets, our PCAP parser can
identify which services are available on the network as well as the IP
addresses of the endpoints and store these in the IR.

\textbf{DHCP}~\cite{rfc2131} is a network bootstrapping protocol that allows
endpoints to exchange configuration, including which network address to use.
Our PCAP parser uses these to create or update endpoints in the IR, based on
hardware address, with their assigned network addresses.

\textbf{ICMP}~\cite{rfc792} is a protocol that is used for network status
messages and diagnostics. Our parser creates new endpoints based on echo and
echo reply messages.

\textbf{TCP SYN}~\cite{rfc793} is the first packet in a TCP connection that
includes many parameters for the connection. Previous work has shown that the
TCP parameters are operating system dependent which can be fingerprinted
(implemented in p0f~\cite{zalewski2014p0f}). We ported p0f to Go and to run on
top of GoPacket so that our PCAP parser could annotate endpoints with its
suspected operating system.

Our PCAP parser combines data from these various protocols to create endpoints
and populate their metadata. There are many inconsistencies that can occur
passively monitoring a live network. Our default behavior is to drop
inconsistent endpoints. In future work, developing a more rigorous approach to
resolving inconsistencies would substantially improve the quality of the
generate IR.

\subsection{minemiter}

\emph{minemiter} is the model generation tool for the discovery IR. It walks
the IR, invoking one or more templates for every node in the graph, and emits a
minimega script that can then be used to boot the model. The core feature of
minemiter is its use of user definable templates (based on Go's template
package\footnote{\url{https://golang.org/pkg/text/template}}) to provide
actions on qualities of endpoints. For example, a node in the IR may have a
network connection on network \emph{63}, and a MAC address of
\emph{de:ad:be:ef:ca:fe}, as is shown in Figure~\ref{fig:host_ir}. From this, a
templated action to generate the necessary minimega command could look like:

\begin{verbatim}
{{ if $e.D.mac }}
vm config net network-{{$e.N}},{{$e.D.mac}}
{{ end }}
\end{verbatim}

This production would generate the valid minimega command:

\begin{verbatim}
vm config net network-63,de:ad:be:ef:ca:fe
\end{verbatim}

Templates are executed in sorted-name order, and minemiter expects template
names to begin with a single letter (A-Z) and a two-digit number, such as
S70network.template. Using this scheme, minemiter will execute every ``A''
template in numeric order against every node in the IR, followed by every ``B''
template, and so on. This enables the template writer to have up to 26 passes
across every node in the IR.

By processing templates in this way, the designer can approach model building
iteratively. For example, a default set of templates will generate an abstract,
Linux container-based experiment with no behavioral model. From this point, the
model designer can write templates specific to their experiment, such as a
template that generates Windows KVM-based endpoints. In this case, the Windows
template would be ordered before the default Linux container templates, and any
endpoints that matched on the template criteria (such as having the metadata
os=windows), would emit a Windows VM configuration. The Windows template would
then indicate not to continue in this pass for this node. If the criteria do
not match, minemiter simply falls through to the next template, in this case
the default Linux container templates. By using this approach, the model
designer can both iterate on successively higher fidelity models without having
to modify the source IR, as well as easily modify the level of abstraction in
the experiment by simply adding or removing key templates.

The templates are so flexible that different sets of templates could even be
used to generate models for different emulation platforms. Alternatively, it
may be simpler to transliterate the discovery IR to whatever IR the other
desired emulation platforms supports.

\section{SCinet 2016}

During Supercomputing 2016, we worked to build a discovery-based model of
SCinet 2016 as NRE participants.

\subsection{Router Configurations}

\begin{figure*}
    \centering
    \includegraphics[width=0.8\textwidth]{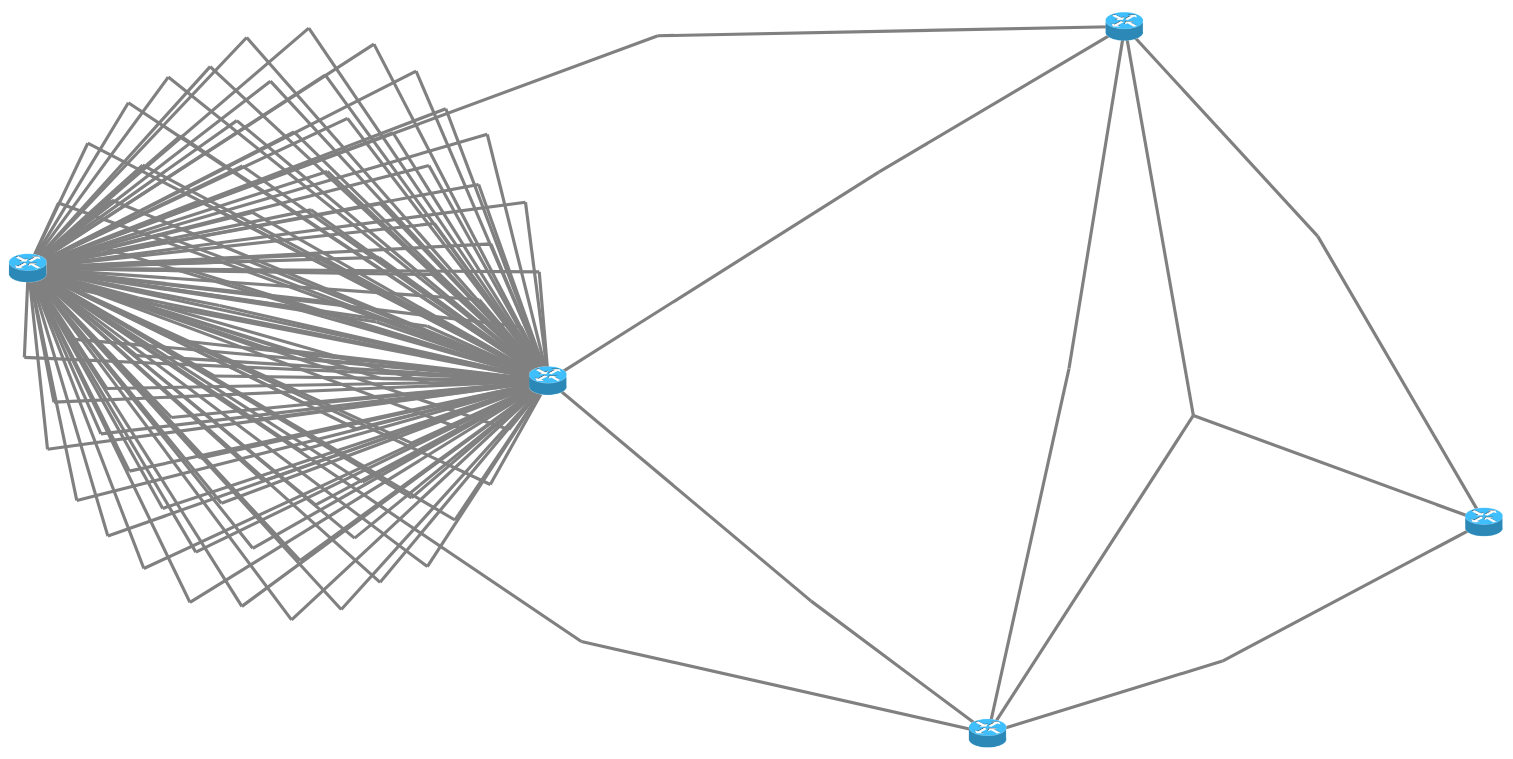}
    \caption{Model based on router configurations that shows the many
    interconnections between the two aggregation routers, their connections to
    the core routers, and finally the connections between the core routers and
    the conference router.}
    \label{fig:sc16-routers}
\end{figure*}

Our first step to modeling SCinet was to parse the router configurations in
order to determine the structure of the network. SCinet 2016 included five
routers from four different vendors: Arista, Brocade, Juniper, and Cisco. The
parser for each configuration file created a new node in the IR and created an
edge for each IP assigned to the router. The parsers inspect the IP address and
subnet assigned to each edge and connect the edges that share the same subnet.
We were only interested in constructing a simple layer-3 model so we ignored
other configurations such as VLAN-tagging for certain interfaces.
Figure~\ref{fig:sc16-routers} shows a picture of the resulting IR.

\subsection{dhcpd Logs}

Once we had the structure of the network, we built a new parser for
\emph{dhcpd} logs to populate the endpoints on the network. We chose to use
dhcpd logs for several reasons. First, they include devices across many of the
conference subnets. Second, these logs are simple to parse. Third, they are
small in comparison to the terabytes of PCAP that we would otherwise have to
process to obtain the same information.

The parser scans the logs looking for DHCP acknowledgements (ACK). ACKs are of
the format:

\begin{verbatim}
DHCPACK on 140.221.X.Y to <MAC> \
    (<HOSTNAME>) via 140.221.X.1
\end{verbatim}

For each ACK, the parser queries the IR to see if it already contains an
endpoint with the given MAC and if it does not, it creates one. It then updates
the endpoint with the leased IP address and hostname. This parser took only a
few hours to implement and run on the logs which demonstrates a strength of the
discovery toolset -- adding new parsers is trivial.

Figure~\ref{fig:sc16-partial} shows a picture of the resulting IR, omitting the
two largest subnets: the wireless network and the classrooms. These subnets
make the graph too large to easily display. To omit these subnets, we simply
wrote a small \emph{trim} tool that walks the IR and marks nodes for deletion
based on a query. We ran this for each subnet, unmarked the nodes representing
routers, and then ran it again to delete the marked nodes.

The dhcpd logs that we processed contained a total of 9,525 endpoints. From the
hostnames they request, we estimate that there were at least 2,100 Android and
1,900 iPhone devices. Another 500 were iPads and another 71 were Windows
devices. Based on this information, we changed the icon in the visualization to
show where the different OSes appear in the network. Upon doing so, we
discovered that one subnet is entirely Windows devices.

\subsection{Emulation}

We took the resulting IR and used minemiter to create a minimega script to boot
a Emulytics model of SCinet 2016. For simplicity, the environment is based on
the default templates -- all endpoints are represented with containers and all
IPs were statically assigned. The script contained over 128k minimega commands
and took just over 14 minutes to run, launching the 8,416 containers across 10
machines. When booting the containers, we discarded endpoints that had the same
IP address or the same hostname. These discarded endpoints suggest improvements
that we can make to our dhcpd parser -- there should be no duplicate IPs for
the time that the DHCP lease is valid and hostnames should be deduplicated by
appending an integer.

\begin{figure}
    \centering
    \includegraphics[width=0.45\textwidth]{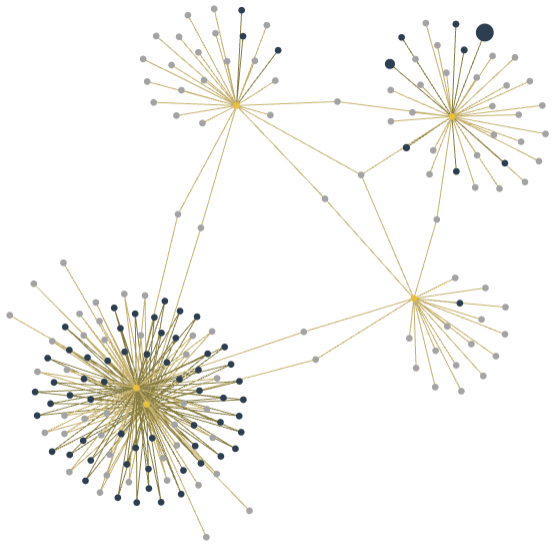}
    \caption{Screenshot of minimega's network graph for SCinet 2016 model}
    \label{fig:sc16-minimega}
\end{figure}

Figure~\ref{fig:sc16-minimega} shows a screenshot of minimega's network graph
which shows how the container networks connect in a running minimega
environment. From this network graph, we can see a similar structure to the
IR-based graphs shown in Figure~\ref{fig:sc16-routers} and
Figure~\ref{fig:sc16-partial}.

\section{Discussion}

We chose the simplest emulation for SCinet 2016: an all container environment
with static IP addresses. There are many ways in which we can improve the
fidelity of the environment using the IR and templates built into the toolset.
However, an open question is how much effort to put into improving the
emulation. For example, we could replace the containers with KVM-based VMs and
launch operating systems matching those of the real devices but that may not
provide any benefit over an all container environment if we are only interested
in studying properties of a firewall or IDS appliance between the registration
network and the WiFi devices.

Furthermore, there is the question of how well our model matches the real
network. So far, we have compared graphs from the Emulytics environment to the
IR-based graphs and the IR-based graphs to the SCinet network diagrams. A more
rigorous analysis of whether the Emulytics model matches the IR model which, in
turn, matches the real network, would be required before claims about the real
network could be made using the Emulytics model. Fortunately, many of the
applications of Emulytics do not require this level of rigor. For example, to
continue with the previous example, studying properties of a firewall or IDS
appliance.

\section{Future Work}

To date, most of our efforts have been towards understanding a static snapshot
of the network we that wish to emulate. In the real-world, networks evolve over
time as when endpoints are added, removed, or simply move to a different
location within the network. This evolution is particularly interesting with
laptops and mobile devices which have high mobility.

The templated model generation technique we employ could be used to perform
transforms on already running experiments. For example, say source network data
from a given network is captured a day apart, and two IRs are created from that
data. The model generator can be used to create a ``base'' model, representing
the first data snapshot. Once the base model is running, the model generator
can use a separate set of templates to determine the difference from the second
model (created from data taken a day later), and create Emulytics platform
commands to modify the booted model. This allows creating temporal snapshots of
experiments and modifying experiments to reflect those temporal changes.
Alternatively, because our IR includes an unstructured map, we could also
annotate devices based on time and change our tools to be aware of time. This
may require significant changes to our tools that currently expect uniqueness
in fields such as MAC address and IP address.

\section{Conclusion}

In conclusion, we have presented our discovery toolset which aims to automate
the map to model process for Emulytics. The toolset contains a daemon that
stores the IR and provides a RESTful API for tools, parsers, and model
generators to communicate with. On the backend, we describe minemiter, a tool
that takes a template-based approach to generate commands to instantiate a
model of the network in an emulation platform. We have applied our tools and
techniques to model SCinet 2016 which we booted in minimega.

\section*{Acknowledgements}

Sandia National Laboratories is a multimission laboratory managed and operated
by National Technology \& Engineering Solutions of Sandia, LLC, a wholly owned
subsidiary of Honeywell International Inc., for the U.S. Department of Energy’s
National Nuclear Security Administration under contract DE-NA0003525.

This paper describes objective technical results and analysis. Any subjective
views or opinions that might be expressed in the paper do not necessarily
represent the views of the U.S. Department of Energy or the United States
Government.

\bibliographystyle{plain}
\bibliography{main}

\begin{thebibliography}{10}

\bibitem{bellard2005qemu}
{\sc Bellard, F.}
\newblock Q{EMU}, a {F}ast and {P}ortable {D}ynamic {T}ranslator.
\newblock In {\em USENIX Annual Technical Conference, FREENIX Track\/} (2005),
  pp.~41--46.

\bibitem{benzel2006experience}
{\sc Benzel, T., Braden, R., Kim, D., Neuman, C., Joseph, A., Sklower, K.,
  Ostrenga, R., and Schwab, S.}
\newblock Experience with {DETER}: A {T}estbed for {S}ecurity {R}esearch.
\newblock In {\em Testbeds and Research Infrastructures for the Development of
  Networks and Communities, 2006. TRIDENTCOM 2006. 2nd International Conference
  on\/} (2006), IEEE, pp.~10--pp.

\bibitem{berthier2010nfsight}
{\sc Berthier, R., Cukier, M., Hiltunen, M., Kormann, D., Vesonder, G., and
  Sheleheda, D.}
\newblock Nfsight: Netflow-based {N}etwork {A}wareness {T}ool.
\newblock In {\em Proceedings of LISA’10: 24th Large Installation System
  Administration Conference\/} (2010), p.~119.

\bibitem{rfc6762}
{\sc Cheshire, S., and Krochmal, M.}
\newblock {Multicast DNS}.
\newblock RFC 6762, Feb. 2013.

\bibitem{de2014using}
{\sc De~Oliveira, R. L.~S., Shinoda, A.~A., Schweitzer, C.~M., and Prete,
  L.~R.}
\newblock Using {M}ininet for {E}mulation and {P}rototyping
  {S}oftware-{D}efined {N}etworks.
\newblock In {\em Communications and Computing (COLCOM), 2014 IEEE Colombian
  Conference on\/} (2014), IEEE, pp.~1--6.

\bibitem{rfc2131}
{\sc Droms, R.}
\newblock {Dynamic Host Configuration Protocol}.
\newblock RFC 2131, Mar. 1997.

\bibitem{filip2010bird}
{\sc Filip, O., Forst, L., Machek, P., Mares, M., and Zajicek, O.}
\newblock B{IRD} {I}nternet {R}outing {D}aemon.
\newblock {\em NANOG-48, Austin, TX\/} (2010).

\bibitem{govindan2000heuristics}
{\sc Govindan, R., and Tangmunarunkit, H.}
\newblock Heuristics for internet map discovery.
\newblock In {\em INFOCOM 2000. Nineteenth Annual Joint Conference of the IEEE
  Computer and Communications Societies. Proceedings. IEEE\/} (2000), vol.~3,
  IEEE, pp.~1371--1380.

\bibitem{habib2008virtualization}
{\sc Habib, I.}
\newblock Virtualization with kvm.
\newblock {\em Linux Journal 2008}, 166 (2008), 8.

\bibitem{hibler2008large}
{\sc Hibler, M., Ricci, R., Stoller, L., Duerig, J., Guruprasad, S., Stack, T.,
  Webb, K., and Lepreau, J.}
\newblock Large-scale {V}irtualization in the {E}mulab {N}etwork {T}estbed.
\newblock In {\em USENIX Annual Technical Conference\/} (2008), pp.~113--128.

\bibitem{jacobson1988traceroute}
{\sc Jacobson, V.}
\newblock Traceroute {S}oftware.
\newblock {\em Lawrence Berkeley Laboratories\/} (1988).

\bibitem{kim2007efficient}
{\sc Kim, S., and Harfoush, K.}
\newblock Efficient estimation of more detailed internet ip maps.
\newblock In {\em Communications, 2007. ICC'07. IEEE International Conference
  on\/} (2007), IEEE, pp.~377--384.

\bibitem{lyon2009nmap}
{\sc Lyon, G.~F.}
\newblock {\em Nmap {N}etwork {S}canning: The {O}fficial {N}map {P}roject
  {G}uide to {N}etwork {D}iscovery and {S}ecurity {S}canning}.
\newblock Insecure, 2009.

\bibitem{mirkovic2018dew}
{\sc Mirkovic, J., Bartlett, G., and Blythe, J.}
\newblock {DEW}: Distributed experiment workflows.
\newblock In {\em 11th {USENIX} Workshop on Cyber Security Experimentation and
  Test ({CSET} 18)\/} (Baltimore, MD, 2018), {USENIX} Association.

\bibitem{pfaff2015design}
{\sc Pfaff, B., Pettit, J., Koponen, T., Jackson, E.~J., Zhou, A., Rajahalme,
  J., Gross, J., Wang, A., Stringer, J., Shelar, P., et~al.}
\newblock The {D}esign and {I}mplementation of {O}pen v{S}witch.
\newblock In {\em NSDI\/} (2015), pp.~117--130.

\bibitem{rfc826}
{\sc Plummer, D.~C.}
\newblock {An Ethernet Address Resolution Protocol: Or Converting Network
  Protocol Addresses to 48.bit Ethernet Address for Transmission on Ethernet
  Hardware}.
\newblock RFC 826, Nov. 1982.

\bibitem{rfc792}
{\sc Postel, J.}
\newblock {Internet Control Message Protocol}.
\newblock RFC 792, Sept. 1981.

\bibitem{rfc793}
{\sc Postel, J.}
\newblock {Transmission Control Protocol}.
\newblock RFC 793, Sept. 1981.

\bibitem{ricci2014introducing}
{\sc Ricci, R., and Eide, E.}
\newblock Introducing {C}loud{L}ab: {S}cientific {I}nfrastructure for
  {A}dvancing {C}loud {A}rchitecturesand {A}pplications.
\newblock {\em ; login: 39}, 6 (2014), 36--38.

\bibitem{spring2002measuring}
{\sc Spring, N., Mahajan, R., and Wetherall, D.}
\newblock Measuring isp topologies with rocketfuel.
\newblock {\em ACM SIGCOMM Computer Communication Review 32}, 4 (2002),
  133--145.

\bibitem{urias2015emulytics}
{\sc Urias, V., Van~Leeuwen, B., Stout, W., and Wright, B.}
\newblock Emulytics at {S}andia {N}ational {L}aboratories.
\newblock {\em MODSIM World 2015\/} (2015).

\bibitem{yin2004visflowconnect}
{\sc Yin, X., Yurcik, W., Treaster, M., Li, Y., and Lakkaraju, K.}
\newblock Vis{F}low{C}onnect: Netflow {V}isualizations of {L}ink
  {R}elationships for {S}ecurity {S}ituational {A}wareness.
\newblock In {\em Proceedings of the 2004 ACM Workshop on Visualization and
  Data Mining for Computer Security\/} (2004), ACM, pp.~26--34.

\bibitem{zalewski2014p0f}
{\sc Zalewski, M.}
\newblock p0f v3 (version 3.08 b), 2014.

\end{thebibliography}

\end{document}